# Complex magnetic phases and thermodynamics of $CuB_2O_4$

C. H. Lai,[1] M.-J. Hsieh,[2] N. Puri,[2] Yiing-Rei Chen[3], J. K. Wang,[3] Y. H. Liang,[1] P.-Y. Chen,[4] Shin-ichiro Yano,[5] C.-W. Wang,[5] C. H. Du,[1,†] C. L. Huang[5,6,‡], and J.-Y. Lin[2,*]

[1]*Department of Physics, Tamkang University, Tamsui, New Taipei City 25137, Taiwan*

[2]*Institute of Physics, National Yang Ming Chiao Tung University, Hsinchu 300093, Taiwan*

[3]*Department of Physics, National Taiwan Normal University, Taiwan 116, Taiwan*

[4]*Department of Physics and Center for Quantum Frontiers of Research & Technology (QFort), National Cheng-Kung University, Tainan 701, Taiwan*

[5]*National Synchrotron Radiation Research Center, Hsinchu 30092, Taiwan*

**Abstract**

The copper metaborate $CuB_2O_4$ (CBO) has been studied within the parametric space of magnetic field ($B$) and temperature ($T$), showing a series of distinct phases, including commensurate and incommensurate magnetic orders, as well as magnetic soliton phases. However, no single study has simultaneously demonstrated all these phases. moreover, the existence of additional phases below 2 K or above 2 T for $\mathbf{B} \perp c$ has been scarcely explored. This work presents the first comprehensive construction of the complex $B$-$T$ phase diagram for $T$ = 0.1 – 25 K and $B$ = 0 - 9 T ($\mathbf{B} \perp c$), utilizing thermodynamic probes, magnetic measurements, and neutron scattering on a single batch of CBO crystals. Our findings elucidate new magnetic phases and regimes. First-principles calculations help to gain a deeper understanding of this complex phase diagram. The emergence of various phases within such a narrow $B$ and $T$ regime is attributed to two primary mechanisms: the delicate competition between ferromagnetic and antiferromagnetic interactions, and the weak exchange and superexchange interactions within and between the Cu(A) and Cu(B) sublattices. This competition also leads to a magnetically frustrated ground state at absolute zero where the total magnetic entropy release from high to low temperatures is only 0.76 $R\ln2$, lower than $R\ln2$ expected for copper spin-½. Using CBO as a model system, the present work elucidates the intricate mechanisms underlying the formation of complex magnetic phases.

Field-driven phase transitions have garnered significant attention in condensed matter physics due to their multifunctional properties and wide-ranging applications [1-3]. These transitions involve not only the modulation of magnetization but also the thermodynamic interplay between non-equilibrium sub-states, encompassing contributions from charge, spin, and lattice degrees of freedom [4-6]. Complex magnetic phases often emerge from competing interactions and the interplay of various mechanisms. A particularly intriguing case is the copper metaborate $CuB_2O_4$ (CBO), which has captivated researchers with its complex magnetic phase diagram and unique characteristics. CBO exhibits a variety of magnetic phases under the influence of magnetic field ($B$) and temperature ($T$). The presence of two distinct Cu(A) and Cu(B) sites suggests that these magnetic phases arise from the competition between different spin-ordered subsystems of magnetic $Cu^{2+}$ ions [7-8]. Previous investigations have revealed several phases in the $B$-$T$ phase diagram for $T < 20$ K, including incommensurate magnetic structures in specific regimes. The sequence of phase transitions observed through magnetic and optical studies ranges from paramagnetic (PM) to a commensurate phase of canting antiferromagnetic order (around $T_N = 20$ K), transitioning into an incommensurate phase ($< T^* = 10$ K) with chiral and elliptical spin structures [9-13]. Additionally, elastic neutron scattering studies have identified a phase with a magnetic order resembling a soliton lattice at low temperatures below 4 K, within a narrow magnetic field range around $1.3 \pm 0.2$ T. Two incommensurate phases have been observed in a magnetic field below 10 K [5], with the propagation vector along the $c$ axis. This makes CBO a system with a rich phase diagram and a series of magnetic phase transitions that are not yet fully understood.

To further advance the understanding of CBO, it is crucial to gain new insights into its complex phase diagram through a more comprehensive thermodynamic study of field-driven phase transitions. In this paper, we construct a richer *B-T* phase diagram than previously known, using magnetic susceptibility, isothermal magnetization, specific heat, and neutron scattering measurements, with $\mathbf{B} \perp$ the *c* axis. The phase diagram can be conceptually understood through the competition between ferromagnetic and antiferromagnetic interactions, and the weak exchange and superexchange interactions within and between the Cu(A) and Cu(B) sublattices.

Single crystals of CBO were synthesized using the flux method [1]. The synthesized crystals are sizable and have a shining blue color, as shown in Figure S1 in supplemental material (SM). For the neutron scattering study, to avoid the neutron absorption effect caused by $^{10}B$, the crystals were grown using enriched $^{11}B_2O_3$ instead of $^{10}B_2O_3$. The high quality of the as-prepared CBO crystal was verified using an in-house X-ray diffractometer. For the scattering measurement, the tetragonal CBO crystal was aligned in the (H H L) scattering plane, where H and L are in reciprocal lattice units (r.l.u.) of $2\pi/a \sim 0.545$ Å$^{-1}$ and $2\pi/c \sim 1.12$ Å$^{-1}$, respectively. The specific heat experiments were carried out on the CBO crystal by using both He4 and dilution refrigerator heat-pulsed thermal relaxation calorimeters. The neutron elastic scattering experiments were carried out at the experimental station SIKA of ANSTO. SIKA is equipped with a triple-axis spectrometer with a cold neutron source. For the neutron experiments below 2 K, the measurements were carried out on the high-flux diffractometer WOMBAT of ANSTO with a wavelength of 2.41 Å. The crystal was loaded into a dilution refrigerator. The magnetic measurements were performed on the superconducting quantum interference device (SQUID) of Quantum Design®.

### I. *C(T) with $T \geq 2$ K and B up to 9 T*

We first show the temperature dependent (2 - 25 K) heat capacity $C(T)$ measurements for a CBO single crystal to investigate the magnetic orderings at zero and in applied magnetic fields up to 9 T, as shown in Fig. 1(a). At zero field, there are two sharp peaks of $C(T)$ at $T_N$ = 20 K and $T^*$ = 9 K denoted as transitions to $C_1$ and $I_1$ phases, respectively, consistent with the transitions observed by the magnetization measurements in Fig. S2 in SM. Therefore, these two peaks of $C(T)$ indicate magnetic orderings previously reported in the literature. The comprehensive $C(T,B)$ data set in Fig. 1(a) reveals interesting observations on $C_1$ and $I_1$ which were not shown before. The high-temperature one is a commensurate antiferromagnetic ordering [7-9]. The $C(T)$ peak significantly shifts to higher temperatures with increasing $B$. That $C_1$ shows such an obvious ferromagnetic characteristic and remains a robust peak in $B$ up to 9 T is interesting. On the contrary, the lower temperature peak is associated with an incommensurate antiferromagnetic ordering [7-9], and $T^*$ is significantly suppressed in $B$ and, while the transition seems still robust in $C(T)$ in $B$ = 1T, is suddenly absent as $B$ > 1T. Near 20 K, $C(T)$ exhibits a λ-shaped peak, characteristic of a second order phase transition. Although the $I_1$ peak in $C(T)$ is very sharp at zero field, no temperature hysteresis in $C(T)$ was observed (data not shown), implying that $I_1$ is also a second order phase transition. In addition to the more known phases of $C_1$ and $I_1$ [7-9], at zero field, there is a hump-like feature in $C(T)$ spanning from 2 K to 5 K. With increasing $B$, this feature evolves into a broad peak starting from $B$ = 1 T and the peak significantly moves to higher temperatures. This anomaly (FI) in $B$ suggests a field-induced ferromagnetism-like crossover, and was largely missing in the literature. As discussed below, the FI is key to understand the rich phases of CBO. As a short summary, Fig.

1(a) shows unprecedented $C(T)$ data both at zero field and in $B$ up to 9 T, and reveals comprehensive evolutions of $C_1$, $I_1$, and FI in $B$. More features in Fig. 1(a) will be discussed in the following.

*II. C(T) below 2 K and the entropy analysis*

To investigate whether new phases exist below 2 K, we extended the specific heat measurements down to 100 mK, as shown in Figure 2(a). At $B = 0$, a peak of $C/T$ around 1 K was observed. We notice that the neutron scattering data hinted at a discontinuity at the same temperature [7,9]. Therefore, the 1-K anomaly is most probably associated with the second-order magnetic phase transition. The magnetic susceptibility measurements also evidence this phase transition (see Fig. S3(a) in SM). We denote this phase below 1 K as $X_2$. When a small magnetic field of 0.2 T is applied, the transition becomes first order, manifested by a delta-function-like peak in $C/T$. The first-order phase transition was also hinted at via resonance absorption spectroscopy in a previous report [14]. From the Landau's theory point of view, this implies that the free energy for the magnetic ground state is better described by an expansion to the sixth or higher order in the order parameter [15]. As the field of 0.75 T is applied, the first-order transition becomes the second-order one again before this phase is suppressed in higher fields.

There existed another feature slightly below 2 K at $B \sim 0$ mostly evidenced by the magnetic measurements in Fig. S3(a), which transition temperature is sensitive to a small applied $B$ and becomes suppressed more quickly than that of $X_2$. We denoted this feature as the $X_1$ phase. In any case, *the present study for the first time demonstrates thermodynamic evidence to nail down the existence of both $X_1$ and $X_2$ phases.*

Additionally, it is found that both phases do not show significant magnetic anisotropy with $B$ along the $a$- or $c$- axis (see Fig. S9.).

The CBO lattice consists of $BO_4$ tetrahedrons with $Cu^{2+}$ ions between them positioned in two inequivalent crystallographic positions. Interactions amongst these two distinct copper sites result in complex magnetic behavior [16-18]. One is $Cu^{2+}$ ion at A site (4*b*) abbreviated as Cu(A) and the other is $Cu^{2+}$ ion at B site (8*d*) abbreviated as Cu(B). Boehm et al. concluded that only one of the two inequivalent $Cu^{2+}$ sublattices (Cu(A)) fully orders down to 1.5 K through neutron scattering measurements [9]. This suggests the 1 K transition in Fig. 2(a) mainly originates from the magnetic ordering of the Cu(B) sublattice. After subtracting the non-magnetic part from the total specific heat, the magnetic contribution plotted as $C_{\mathrm{mag}}/T$ and the entropy $S_{\mathrm{mag}}$ are shown in Fig. 2(b). At 30 K, $S_{\mathrm{mag}}$ only reaches ~0.76 $R\ln2$, which suggests that part of Cu(B) moments remain frustrated down to absolute zero. This result is backed up by isothermal magnetization measurements in Figure S3(b), which shows that the magnetic moment saturates at only 0.7 $\mu_{\mathrm{B}}$/f. u., smaller than 1 $\mu_{\mathrm{B}}$/f. u. of spin ½ $Cu^{2+}$. Furthermore, nearly two thirds of entropy is released above 10 K. This entropy analysis suggests that only Cu(A) is ordered for the $C_1$ commensurate while Cu(B) remains largely disordered probably due to fluctuations. This scenario is consistent with the large magnetic moment of Cu(A) and a very small moment of Cu(B) in $C_1$ observed through the neutron scattering experiments [9]. Only in the low $T$ phases, Cu(B) begins to substantially participate in magnetic ordering and form a magnetic lattice together with Cu(A). The relation between interactions amongst these two distinct copper sites will be further elucidated later in the section of *Mechanisms for the complicated phases*.

*III. The phases mainly according to magnetic measurements*

The phases surrounded by FI, $C_1$, $I_1$, and the lowest temperature phases $X_1$, $X_2$ are as complicated and intriguing. For example, the temperature-dependent magnetic moment in magnetic fields of 0.5, 1, 1.3, and 1.4 T was depicted in Figure 1 (b). The peak features in each magnetic field correspond to either of the incommensurate phases $I_1$ and $I_2$ indicated with arrows in Figure 1(b). (Here we adapt the $I_2$ phase as in Ref. [19].) Isothermal magnetization curves also show the incommensurate phase in the temperature range of 6 K < $T$ < 10 K. A kink in $M(B)$ indicates a transition that is shifted to higher fields with decreasing temperature, as shown in Fig. S4(a) in SM. Interestingly, the kink transition gets sharper at $T$ = 4, 3, and 2 K, and shows a hysteresis behavior confirming the first-order field-driven phase transition, as shown in Fig. S4(b) and the inset in SM. Combined with the hysteresis *M-H* data in Ref. [19], the present work established that the dashed green phase boundary and the top solid green phase boundary in Fig. 4, both surrounding $I_2$, are of the first order phase transition. In addition, $C(T)$ (inset of Figure 1a) was measured in two particular fields of 1.37 T and 1.45 T. At $B$ = 1.45 T, $C(T)$ shows a sharp peak at 4.1 K attributed to the incommensurate $I_2$ phase. As $B$ is increased to 1.45 T, the peak appears, albeit smaller, at around 3.4 K. These $C(T)$ results evidence that $I_2$, previously discovered by the magnetic measurements [19], is also a thermodynamic phase.

*IV. Neutron scattering data*

Figures 3(a)-(e) show contour plots scanned through the reflection $(1, 1, 0 + \delta)$ as a function of $B$ at $T$ = 8, 6, 4, 3, and 2 K. The reflection intensity at $T$ = 8 and 6 K

gradually decays and eventually disappears at 0.3 T and 1.2 T, demonstrating the significant change of the $q$-wavevector by $B$ [6]. At $T$ = 4 K, the $I_2$ phase is suppressed by a field of 1.36 T with the simultaneous emergence of a distinct incommensurate phase ($I_3$) at 1.32 T. This demonstrates that the translational symmetry of $I_2$ is broken by $B$, and *both $I_2$ and $I_3$ coexist in a narrow field regime of 0.04 T*. Such a symmetry breaking by $B$ could also be observed at $T$= 2 and 3 K. *More importantly, a metastable state was observed between $I_1$ and $I_2$ phases at $T$= 3 and 2 K* [5]. Figure 3(f) shows a linear scan along (1 1 L), demonstrating the coexistence of $I_2$ ($q_2$), the metastable phase ($q_s$), and $I_3$ ($q_3$) with the satellite reflections of 0.1473, 0.1387, and 0.1277, respectively. This coexistence was not reported in the literature. Fig. 3(g) shows the evolution of the incommensurability $\delta_L$ of $I_1$ ($q_1$), $I_2$ ($q_2$), and $I_3$ ($q_3$) as a function of $B$. It shows that the periodicity (proportional to $1/\delta_L$) of $q_2$ increases and of $q_3$ decreases with increasing $B$ in Fig. 3(g). It is also worth noting that the commensurate phase reflection (0 0 2) was reported to revive at $T$= 4.2 K with $B$ = 1.5 T, suggesting a field-driven reentrant transition for the original commensurate phase which exists between 9.5 to 20 K [5]. Further, the neutron elastic scattering is carried out along the $L$-direction of (1 1 0) in the reciprocal space (Fig. 3(g)) showing evidence of various phase transitions induced by changing $T$, as explained in SM.

In addition, the field-dependent integrated intensity of $q_1$, $q_2$, and $q_3$ extracted from the neutron scattering experiments performed at $T$ = 8, 4, 3, and 2 K for both increasing and decreasing $B$ is summarized in Figure S6 in SM. The field-driven phase transition shows a very different behavior at $T$ = 4, 3, and 2 K. At $T$ = 4 K, for increasing field measurement, the reflection intensity remains until $B$ = 1.3 T (Fig. 4); then drops quickly and disappears at about $B$ = 1.34 T. Meanwhile, a second satellite reflection, marked as $q_3$ phase, emerges at $B$ = 1.32 T and its intensity quickly saturates at 1.35 T. For decreasing $B$, the variation of the reflection intensity displays a similar behavior to that for increasing $B$ with a hysteresis width of about 0.1 T. This is attributed to the first-order phase transition. The evolution of the $q_2$ phase as a function of $B$ at $T$ = 3 and 2 K (Figs. 4(c) and (d) shows the similar first-order phase transition as at $T$ = 4 K, but with different $B$ and hysteresis width and with an additional field-induced metastable phase marked as $q_s$. The change of the transition type from the second order at 8 K and 6 K to the first order at and below 4 K involves a different coupling, which could be from the long-range ordered Cu(B).

We further performed neutron diffraction below 2 K to search for additional phases. At zero field, a full diffraction pattern collected at 1.8 K shows pronounced incommensurate satellites at (3,3,0 ± δ), (5,5,0 ± δ), (4,4,2 ± δ) and so on, consistent with Fig. S6(b). At 25 mK, albeit with this incommensurate phase persisting, *additional satellites emerge* at ($H$ ± 0.5, $K$ ± 0.5, $L$), highlighted by magenta circles in Fig. 5b. This demonstrates a new low-temperature magnetic phase $X_2$. The data further indicate that X2 has an incommensurate magnetic order and may coexist with the $I_2$ phase. For a clearer view, we convert the pattern onto the ($q_x$, $q_y$) plane (Fig. S8), where four satellites are resolved around (1.5, 1.5, −1). Because CBO is tetragonal at low temperature, these satellites group into two symmetry-related sets with propagation vectors $k_1$ =$(\varepsilon\ 0\ \delta)$ and $k_2 = (0\ \varepsilon\ \delta)$, $\varepsilon\sim$ 0.065 and $\delta\sim$0.056 in the unit of reciprocal lattice space. Owing to refrigerator thermal-stability issues, experiments near 1 K were not feasible. The $I_2$ to $X_1$ and $X_1$ to $X_2$ transitions, as well as the magnetic structures of $X_1$and $X_2$ at zero and finite fields, warrant further investigation by neutron scattering.

## *V. Phase diagram*

We summarize the present data of magnetization, specific heat, and neutron scattering to construct a *B*-*T* phase diagram *for **B** ⊥c and above 1000 gauss* in Fig. 4. The characteristic second-order phase transition for CBO in *B* ranging from 0-9 T at around 21 K is indicated by a solid orange phase boundary showing the transition from a PM to weak ferromagnetic commensurate state ($C_1$). The solid green line separating $I_1$ and $I_2$ phases is from the present work. The extended green dashed contour is adapted from earlier studies, and ends at a critical point around $T \sim 6$ K and $B \sim 0.6$ T [5,19]. The $I_3$ phase coexists with a commensurate phase, probably $C_1$. Also, in the boundary between the incommensurate phases $I_2$ and $I_3$ at a temperature below 3 K, a small pocket (the shaded narrow region) is noticed, which is referred to as the magnetic soliton phase (S) [5,20]. The evidence of the soliton phase could also be seen in the neutron diffraction measurements Figs. 3(d) and (e). Phases including PM, commensurate $C_1$, incommensurate $I_1$, $I_2$, and $I_3$ phases, and the magnetic soliton phase (S) are consistent with those reported separately in previous works [5,20,21]. We further identify the $X_1$, $X_2$ and FI phases. This is the first time the complete *B*-*T* phase diagram for **B** ⊥c and above 1000 gauss has been identified through comprehensive thermodynamic probes and neutron scattering measurements on one batch of CBO crystals.

## *VI. Mechanisms for the complicated phases*

The data clearly show that, below 10 K and $B \leq 2$ T, the *B*-*T* diagram becomes highly complex. The richness in Fig. 4(b) is especially rare. There have been cases of antiferromagnetic orderings with $T_N$ similar to that of CBO which B has a very smaller effect on the magnetic ordering even up to 9 T. [22,23] There are other cases of complex magnetic phases [6]; however, *it is exceptional to see several magnetic orderings within the span of ~ 1000 gauss and ~ 1 K except in some skyrmions*. [24] We argue that this rich phase diagram is partially due to the merging of two competing interactions. The CBO crystal shows three different incommensurate phases along with a coexistence region of commensurate and incommensurate phases. It is interesting to understand why various phases coexist within such a narrow phase regime. Phenomenologically in Fig. 4(a), the ferromagnetic dashed purple boundary intersects the antiferromagnetic green boundaries around the regime with rich phases. It is plausible that the merging of these two competing interactions leads to a number of quasi-stable states with comparable energy in such a narrow regime.

Results from calculations shed further light on the formation of these complicated phases. Our ab initio work shows two groups of extremely flat conduction bands separated by ~ 1 eV in the frontier zone. The lower energy one, with a $\lesssim$ 70 meV energy span, is comprised of $d_{xy}$ and $d_{x^2-y^2}$ from Cu(A), and $p_x$ and $p_y$ from O(1). The upper one, with a larger energy span of ~ 160 meV, is mainly comprised of $d_{x^2-y^2}$ and $d_{z^2}$, with small components of $d_{yz}$ and $d_{zx}$, all from Cu(B) and also $p$ orbitals from O(2)~O(4). The charge distributions integrated for the two narrow energy zones show clear patterns due to the distinct orientations of the Cu(A) and Cu(B) plaquettes. Unlike the Cu(A) plaquettes which are almost parallel to the *xy*-plane, Cu(B) plaquettes are much more vertical. These Cu flat bands qualitatively explain sharp lines in the second harmonics spectra in Ref. [25]. It is noted that the sharp optical absorption lines in CBO was attributed to the flat phono band. [26] Most importantly, the flatness implies that any exchange or superexchange interaction between Cu(A) and Cu(B) sublattices must be weak. Consequently, a small perturbation via $B$ or $T$ can significantly impact the magnetic orderings. The calculation results also qualitatively explain why the ordering of Cu(B) is weaker than that of Cu(A). Details of the *ab initio* work can be found in SM.

To conclude, our findings reveal new magnetic phases and regimes, and significantly enhance the understanding of the intricate mechanisms for the phase diagram. The emergence of various phases within such a narrow $B$ and $T$ regime is attributed to the delicate competition between ferromagnetic and antiferromagnetic interactions, as well as the weak superexchange interactions within and between the Cu(A) and Cu(B) sublattices. This competition results in a magnetically frustrated ground state at absolute zero, as suggested by the quantitative entropy analysis. The analysis also reveals the distinct magnetic orders of two copper sites. More than half of entropy is released at the incommensurate to commensurate transition. Thus, in the commensurate phase, Cu(B) sublattice highly fluctuates magnetically and only Cu(A) sublattice is ordered. This study not only advances the understanding of field-driven phase transitions in CBO but also contributes to the broader field by elucidating the intricate mechanisms underlying the formation of complex magnetic phases.

**Acknowledgement**

We thank C.-C. Yang at SQUID VSM Lab, Instrumentation Center, National Cheng Kung University (NCKU) for technical support. The authors also gratefully acknowledge the use of MPMS in Instrumentation Resource Center of National Yang Ming Chiao Tung University. CHD is grateful to NSRRC for providing the travel grant to carry out neutron experiments at ANSTO and to ACNS staff for their assistance during the neutron experiments. This paper is supported by the National Science and Technology Council in Taiwan through Grants Nos. NSTC 109-2112-M-006-026-MY3, 109-2112-M-009 -012 -MY3, 109-2112-M-032-012, 112-2124-M-006-011, 112-2112-A49-027.

†chd@mail.tku.edu.tw; ‡clh@phys.ncku.edu.tw; *ago@nycu.edu.tw

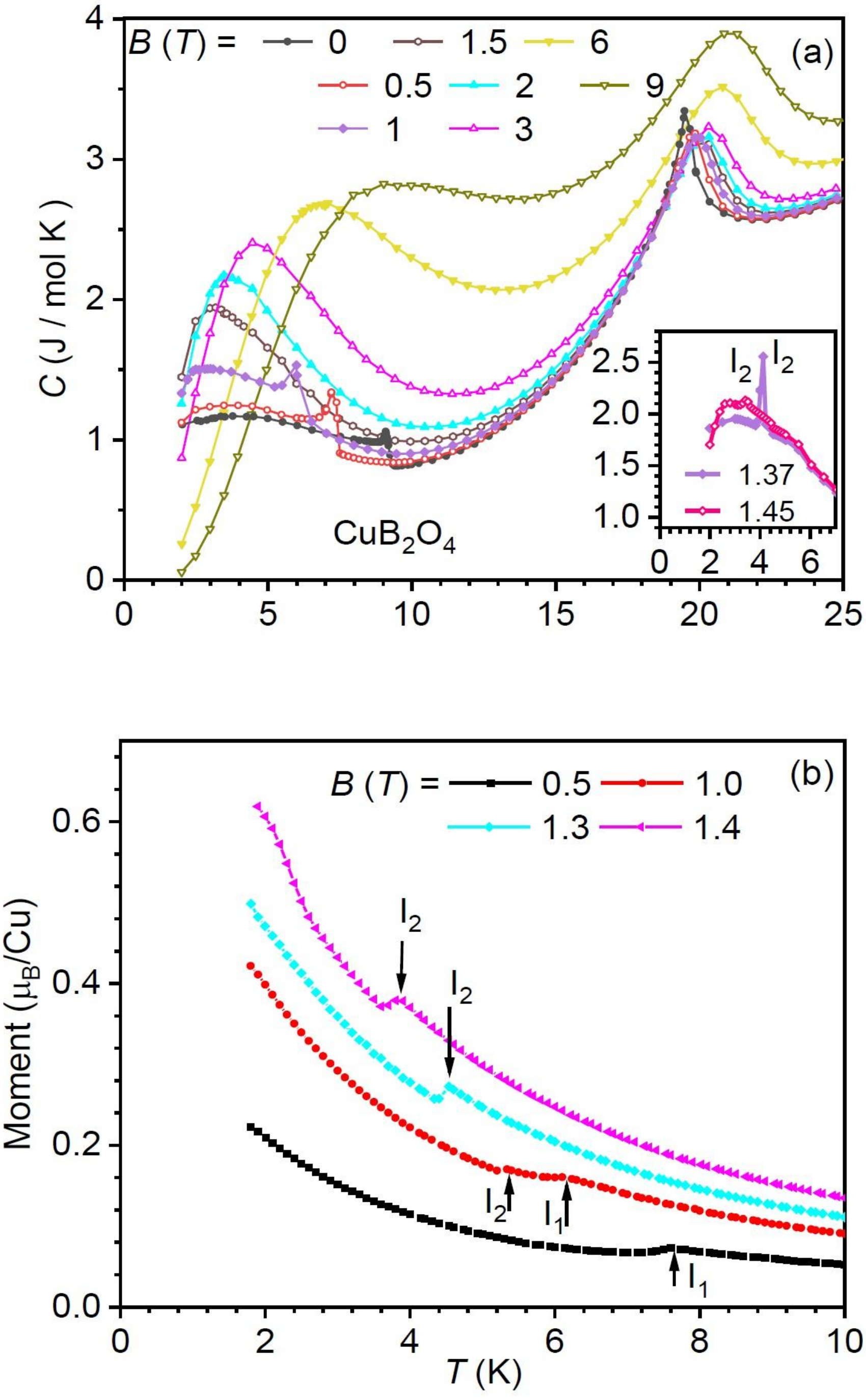

Fig. 1 (a) $C(T)$ curve for $CuB_2O_4$ single crystal in applied magnetic fields $B \perp c$ axis ; at zero field $C(T)$ shows two magnetic ordering temperatures at 10 K ($T^*$) and 20 K ($T_N$), respectively. Inset: $C(T)$ at $B$ = 1.37 T and 1.45 T. (b) Temperature dependence of the magnetic moment in $B$=0.5, 1, 1.3, and 1.4 T.

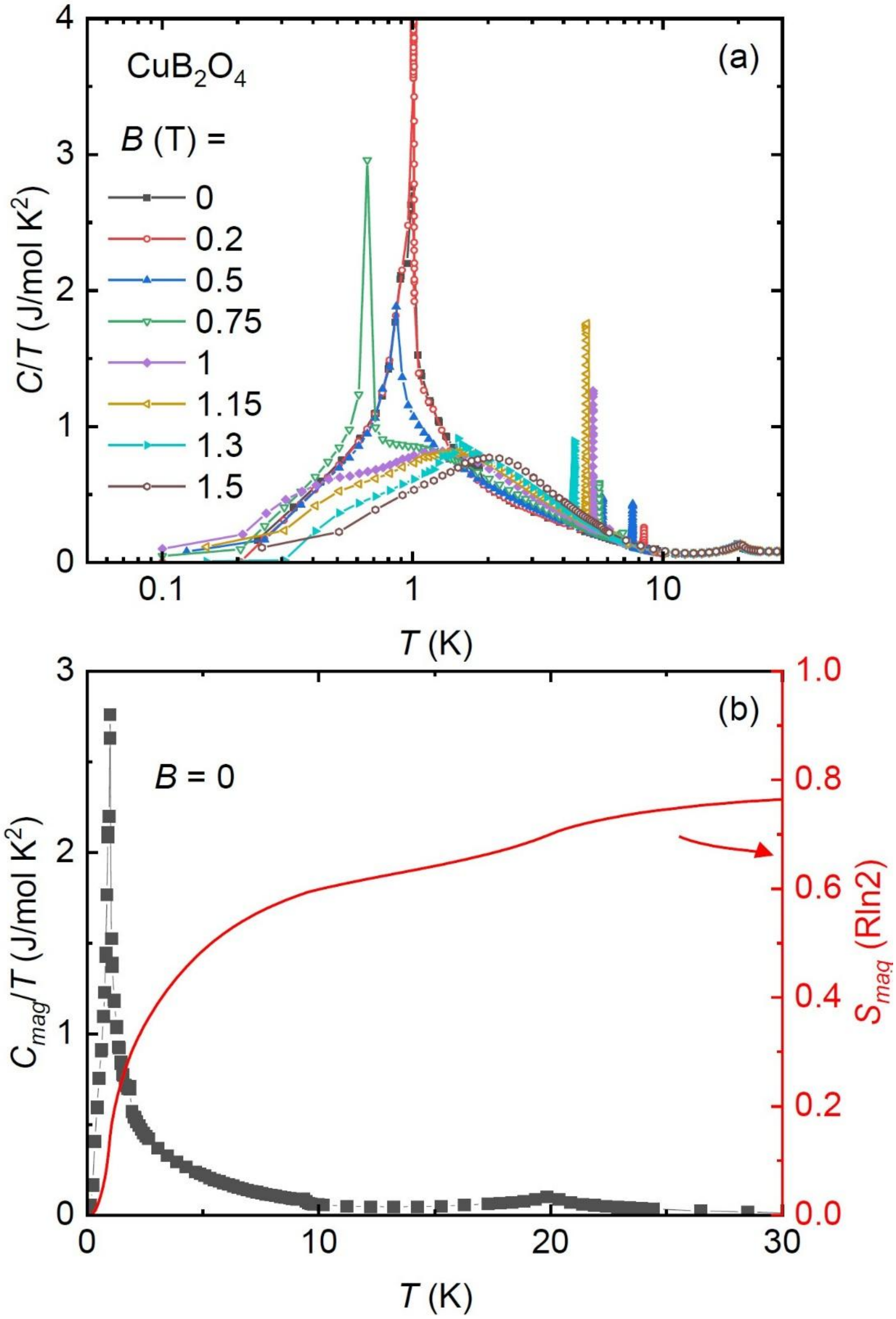

Fig. 2 (a) $C(T)$ down to $T$ = 100 mK for $CuB_2O_4$ single crystal in applied magnetic fields $B \perp c$ axis. (b) Left: magnetic contribution to the specific heat at zero field. Right: magnetic entropy.

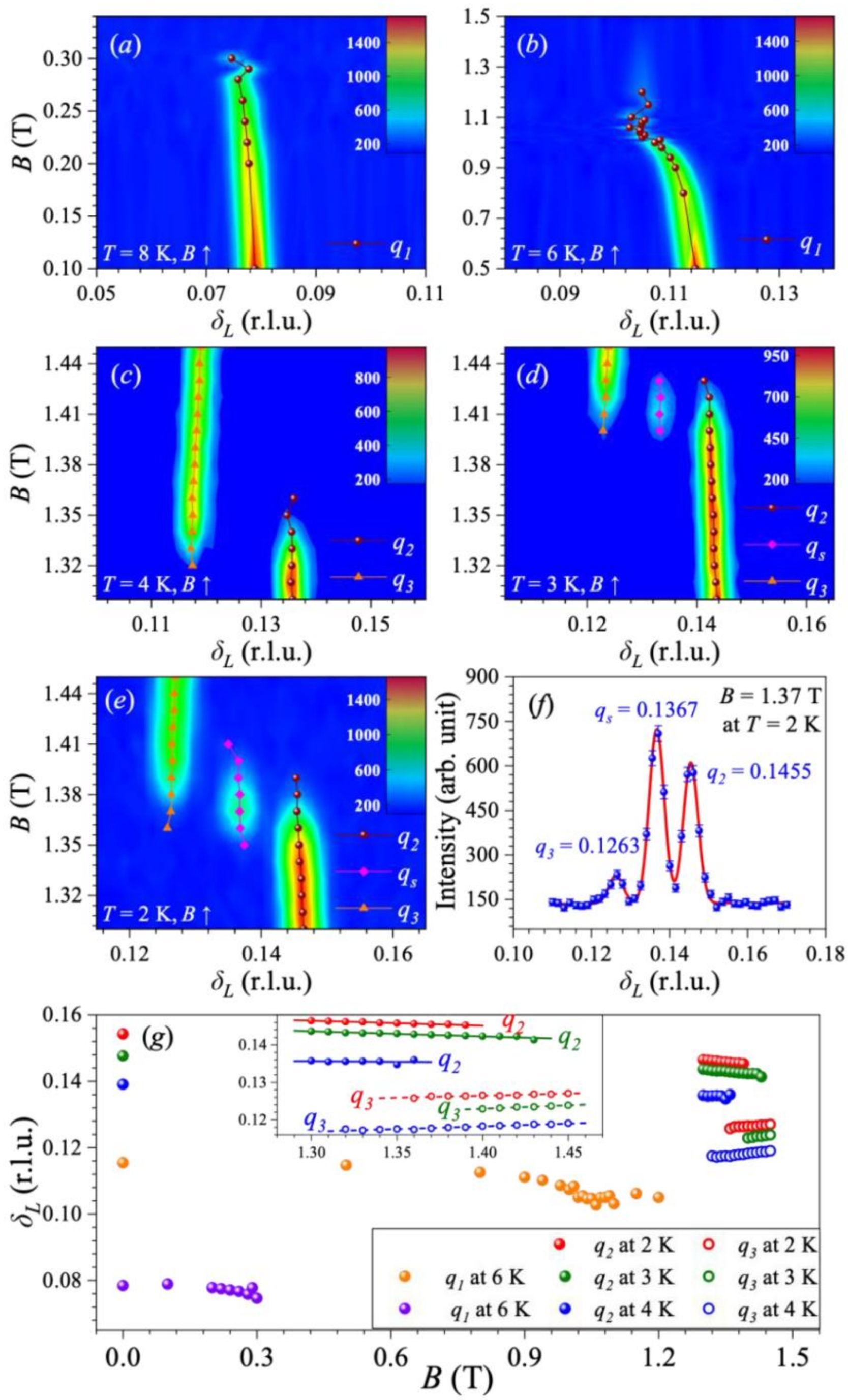

Fig. 3 a-e) Neutron scattering measurements under an application of magnetic field perpendicular to (0 0 1)-direction at *T*= (a) 8, (b) 6, (c) 4, (d) 3, and (e) 2 K. (f) Plot of linear scan along (1 1 L) at *T*=2 K and *B*=1.37 T. (g) Plot of incommensurate *q*-wavevector as a function of $B \perp c$ axis.

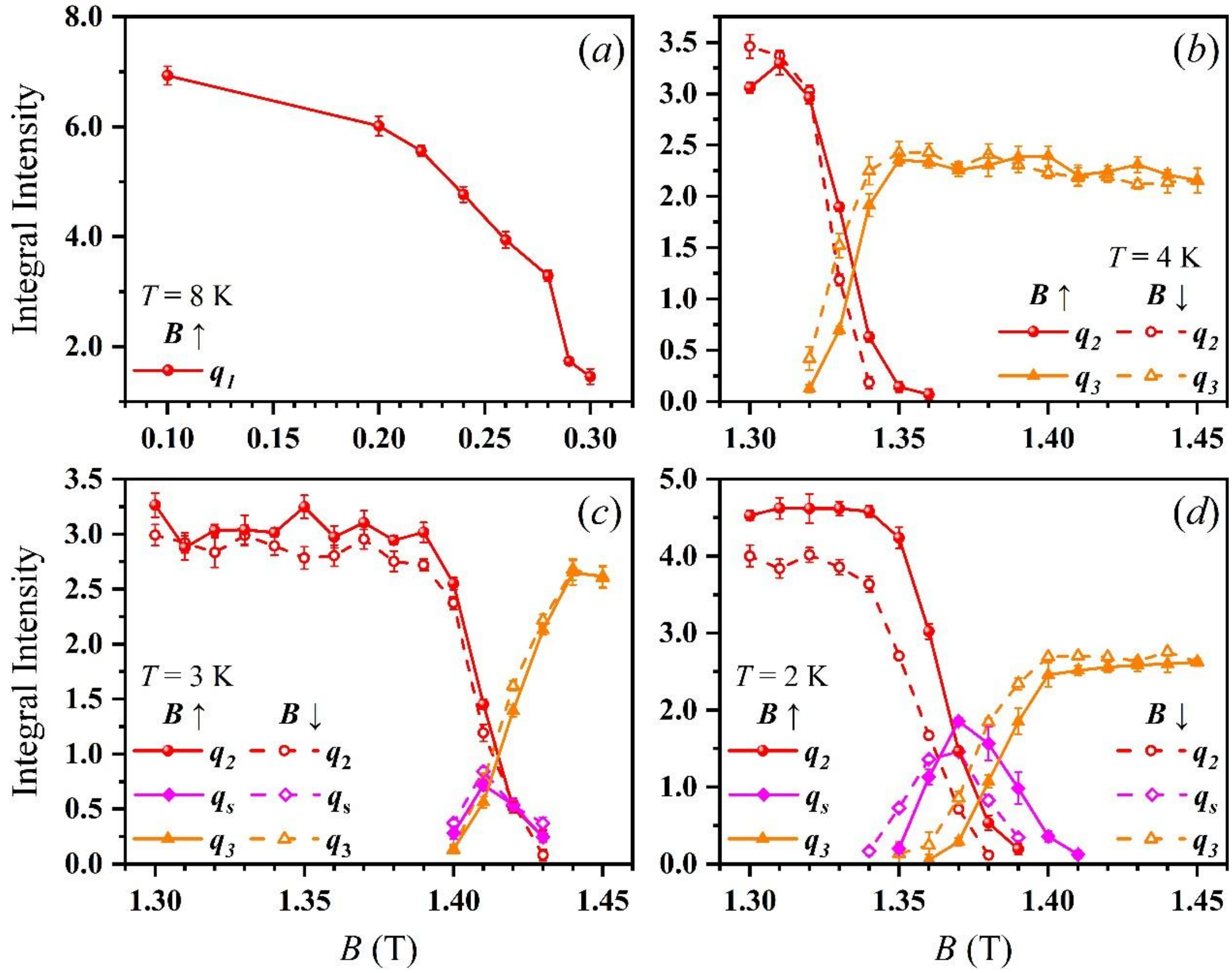

Fig. 4: Integrated intensity of the magnetic reflections of $q_1$, $q_2$, and $q_3$ as a function of magnetic field at $T$ = (a) 8 K, (b) 4 K, (c) 3 K, and (d) 2 K. $B$ ↑ indicates the increasing field measurements and $B$ ↓ the decreasing field.

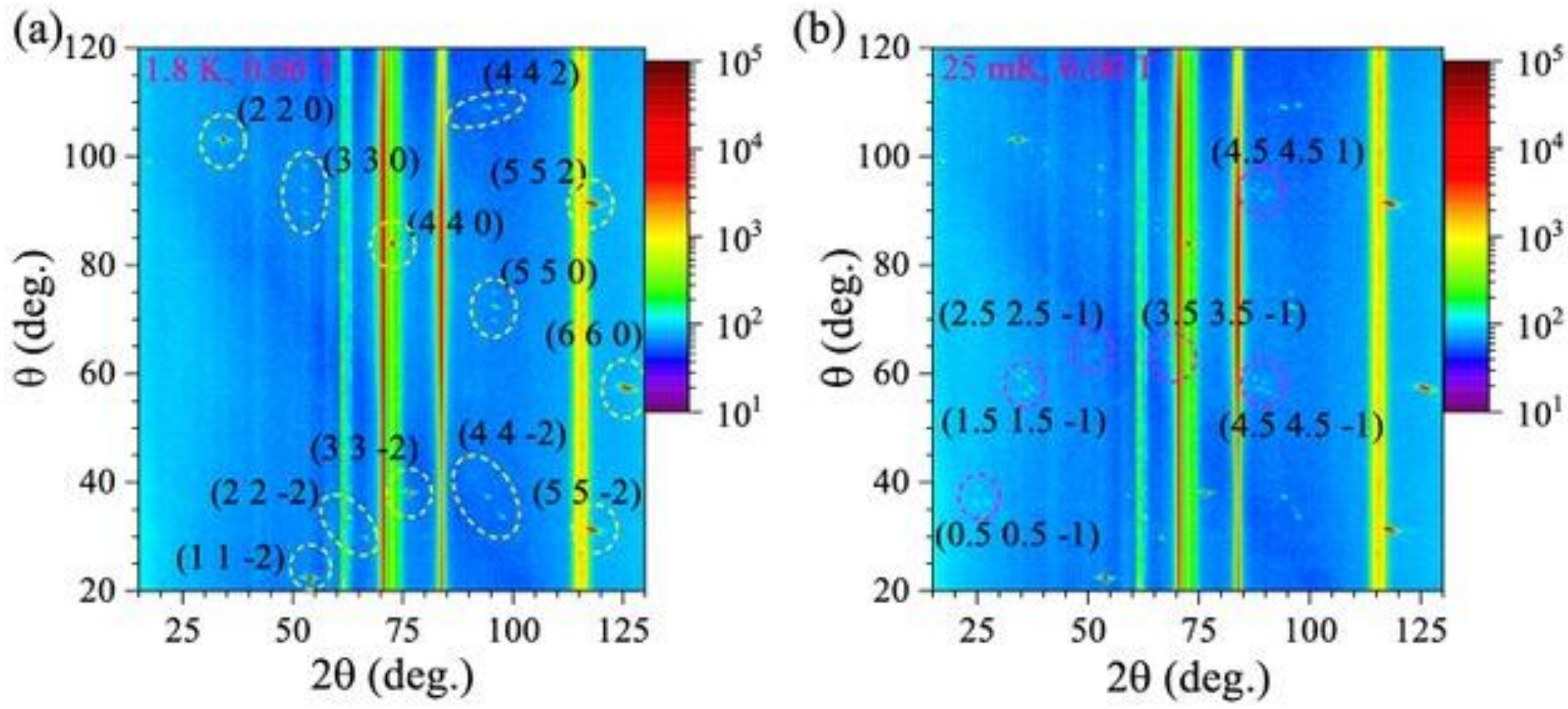

Fig. 5 The incommensurate satellites peaks at (a) 1.8 K and (b) 25 mK. Please note the emergent peaks at 25 mK marked by magenta circles. The incommensurate peaks center around Bragg reflections (such as (2 2 0), (3 3 0), and so on) at 1.8 K and are marked by the yellow circles.

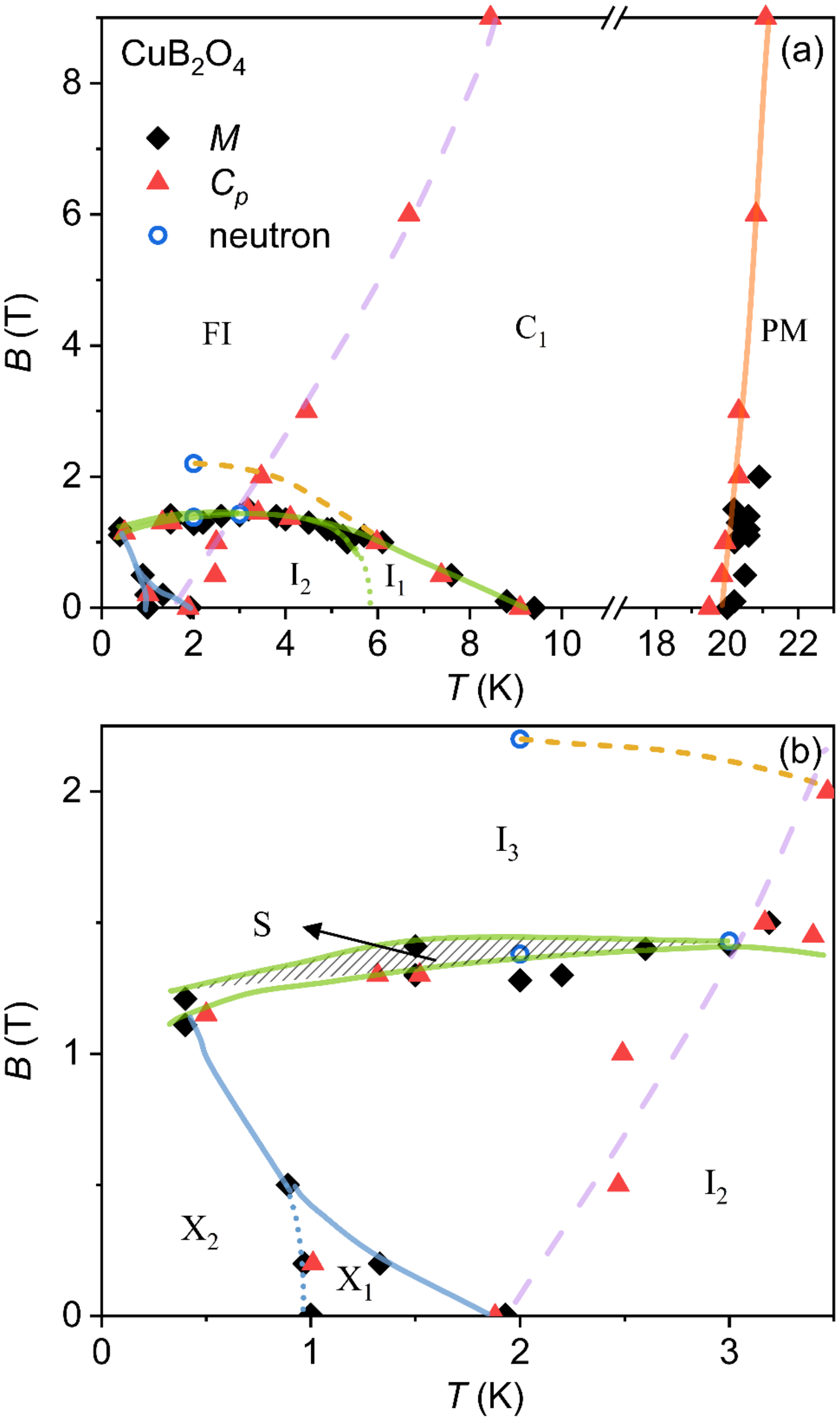

Fig. 6 The magnetic phase diagram of $CuB_2O_4$ in temperature range of (a) 2-25 K and (b) 0-3.5 K with ⊥ *c* axis.

**Supplementary materials**

# Complex magnetic phases and thermodynamics of $CuB_2O_4$

C. H. Lai,[1] M.-J. Hsieh,[2] N. Puri,[2] Yiing-Rei Chen[3], J. K. Wang,[3] Y. H. Liang,[1] P.-Y. Chen,[4] Shin-ichiro Yano,[5] C.-W. Wang,[5] C. H. Du,[1,†] C. L. Huang[5,6,‡], and J.-Y. Lin[2,*]

[1]*Department of Physics, Tamkang University, Tamsui, New Taipei City 25137, Taiwan*

[2]*Institute of Physics, National Yang Ming Chiao Tung University, Hsinchu 300093, Taiwan*

[3]*Department of Physics, National Taiwan Normal University, Taiwan 116, Taiwan*

[4]*Department of Physics and Center for Quantum Frontiers of Research & Technology (QFort), National Cheng-Kung University, Tainan 701, Taiwan*

[5]*National Synchrotron Radiation Research Center, Hsinchu 30092, Taiwan*

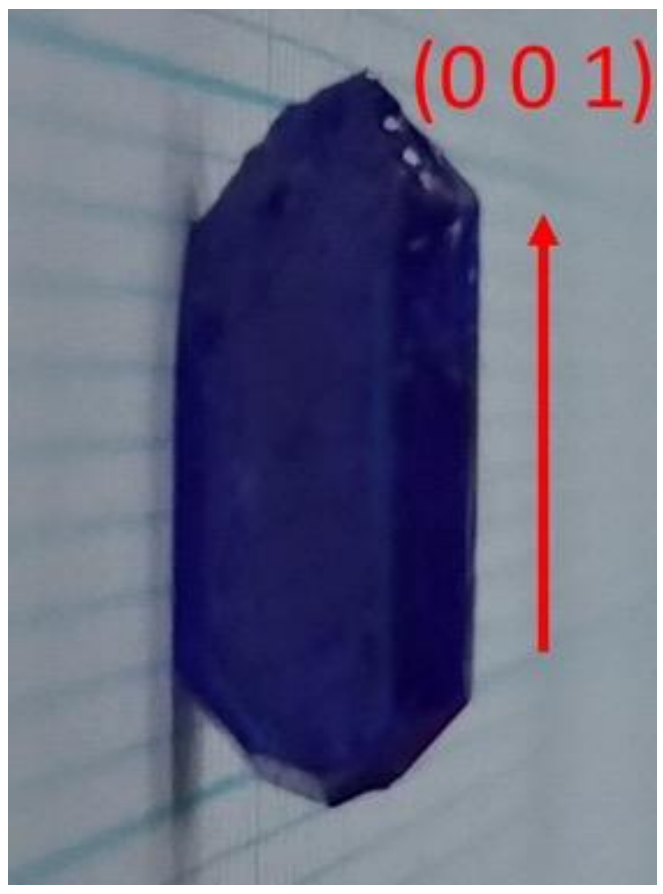

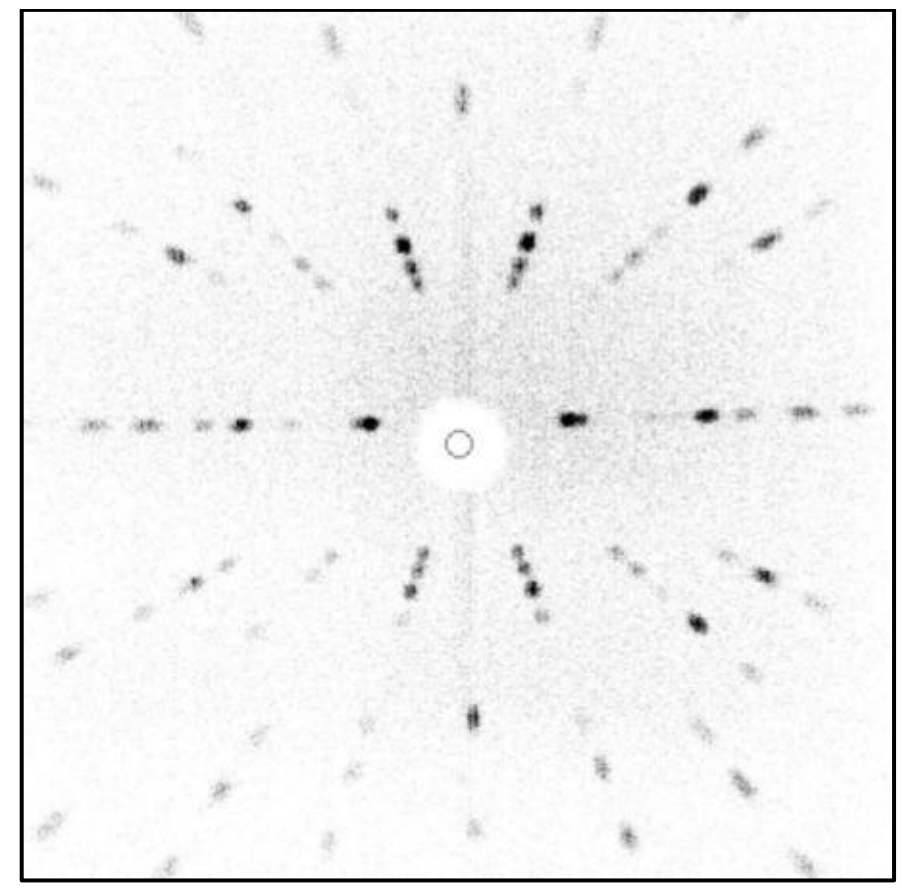

Figure S1: The as-synthesized CBO single crystal.

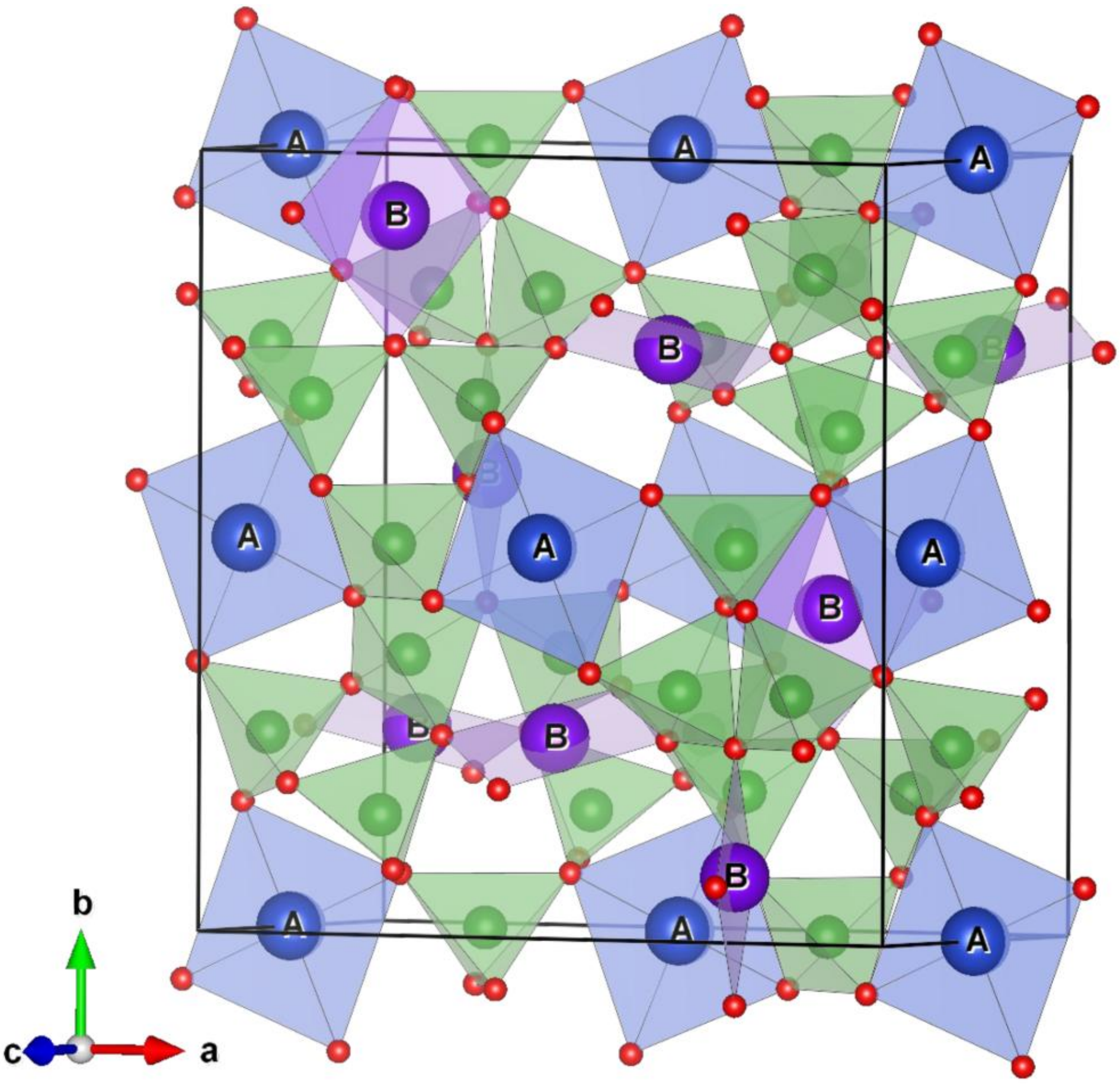

Figure S2: The crystal structure of CBO. The lattice is tetragonal with the lattice constants $a$ = 11.565 Å and $c$ = 5.620 Å. The blue balls denotes Cu(A); purple balls as Cu(B); green balls as B; red balls as O.

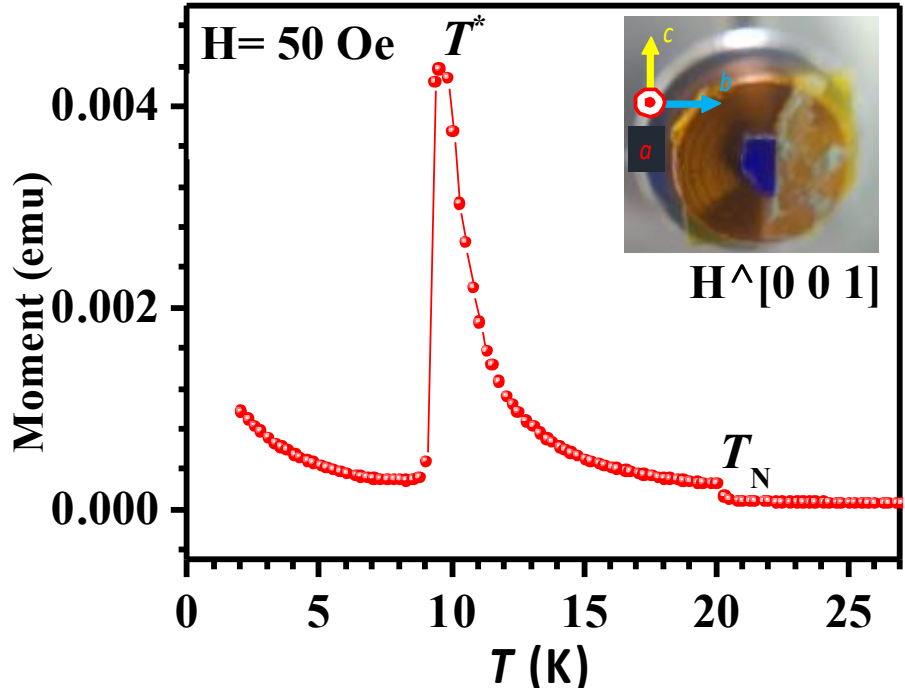

Figure S3: Magnetization measurements of CBO crystal.

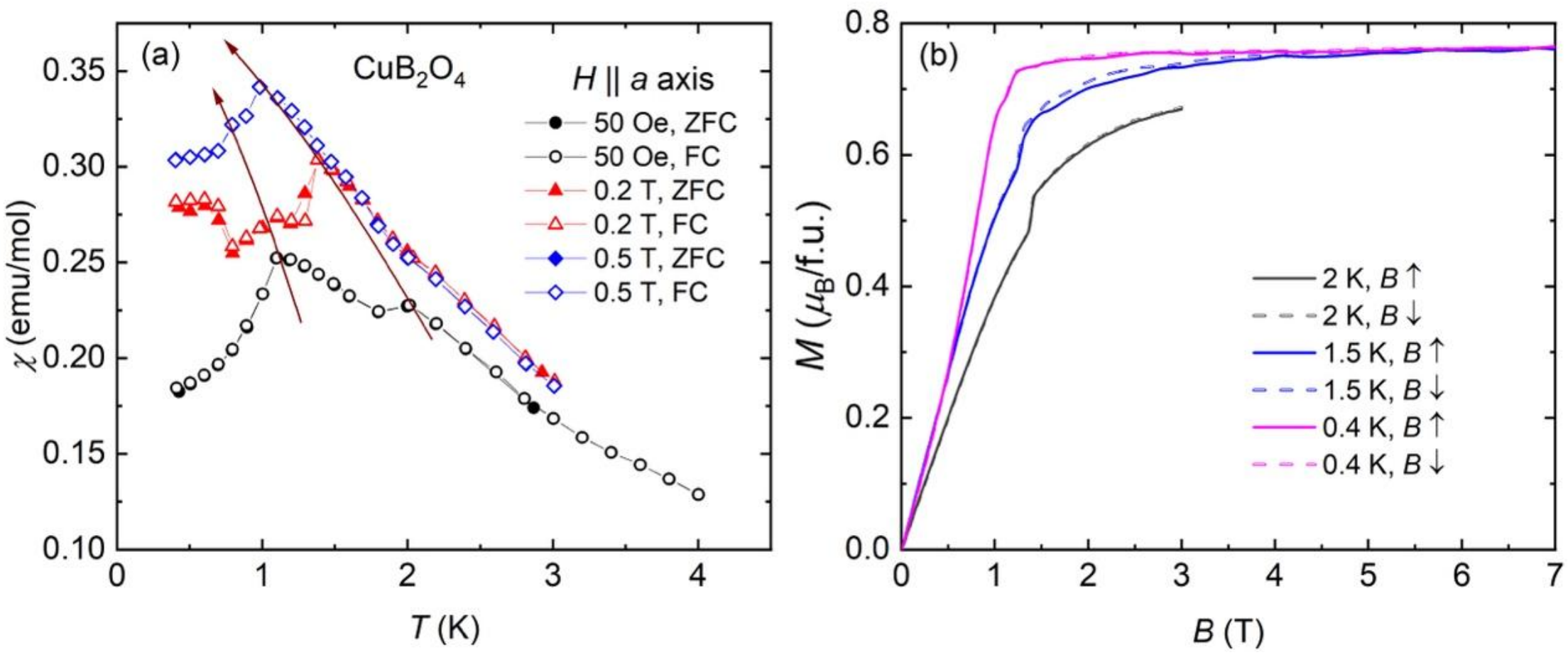

Figure S4: Temperature dependence of magnetic susceptibility measurements (a) and isothermal magnetization curves (b) of CBO crystals.

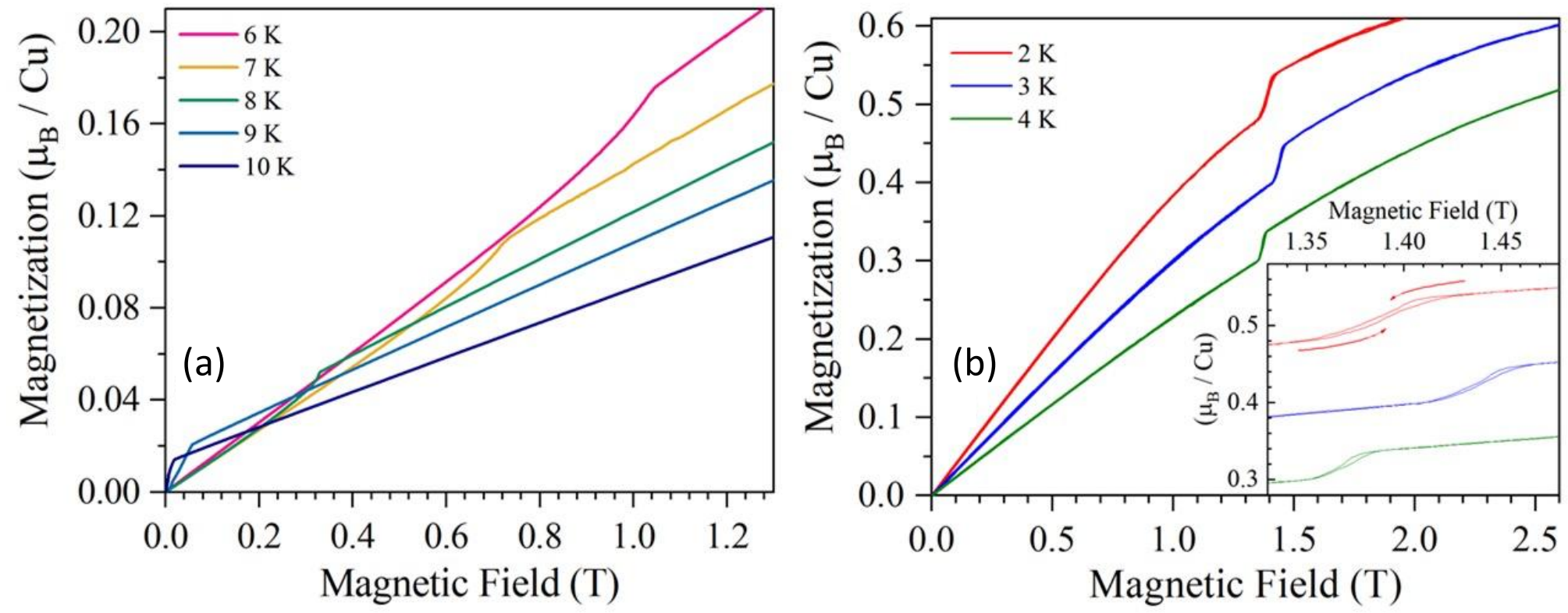

Figure S5: Temperature dependences of the magnetic moment at different magnetic fields ranging from $B$ = 0.05 to 2 T.

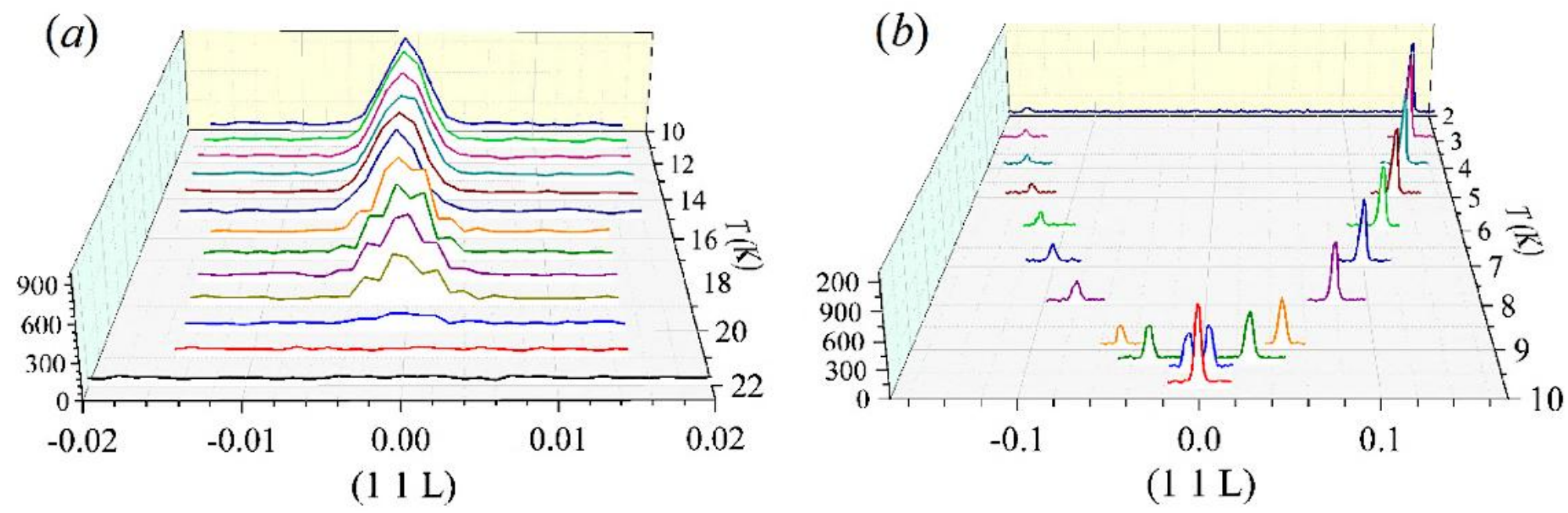

Figure S6: **Neutron scattering data as a function of temperature.** Measurements were conducted in two temperature regions: (a) 10 K ≤ $T$ ≤ 22 K and (b) 2 K ≤ $T$ ≤ 10 K, along the L-direction of the position (1 1 L). During the cooling process, the (1 1 0) reflection intensities begin to increase from $T$ = 20 K with relatively no change in peak position (Fig. 3(g)), indicating the onset of the antiferromagnetic phase with a propagation vector $\vec{q} = 0$ [1]. Upon further cooling, the reflection starts to split at approximately $T$ ~ 9 K, with a propagation wavevector $\vec{q} = (1, 1, 0 \pm \delta)$, where the incommensurability $\delta$ is temperature-dependent. The intensity of the reflection (1,1,0+$\delta$) gradually increases, while (1,1,0-$\delta$) decreases as the temperature decreases. This behavior suggests that the periodicity of the incommensurate phase changes from $c$/0.1543 ~ 36 Å at 2 K to c/0.045 ~ 125 Å at 9 K (with c = 5.607 Å at room temperature). This phase transition is identified as the incommensurate phase $I_1$ in this study and is also known as chiral spin ordering. It is noteworthy that previous neutron scattering studies without the application of a magnetic field [7] reported the existence of higher-order satellite reflections between the commensurate-incommensurate transition boundary, providing evidence for the presence of a magnetic soliton lattice. However, in our observations, we did not observe higher-order magnetic reflections within this transition boundary. The absence of third-order reflections might be attributed to the weaker intensity of the first-order reflections.

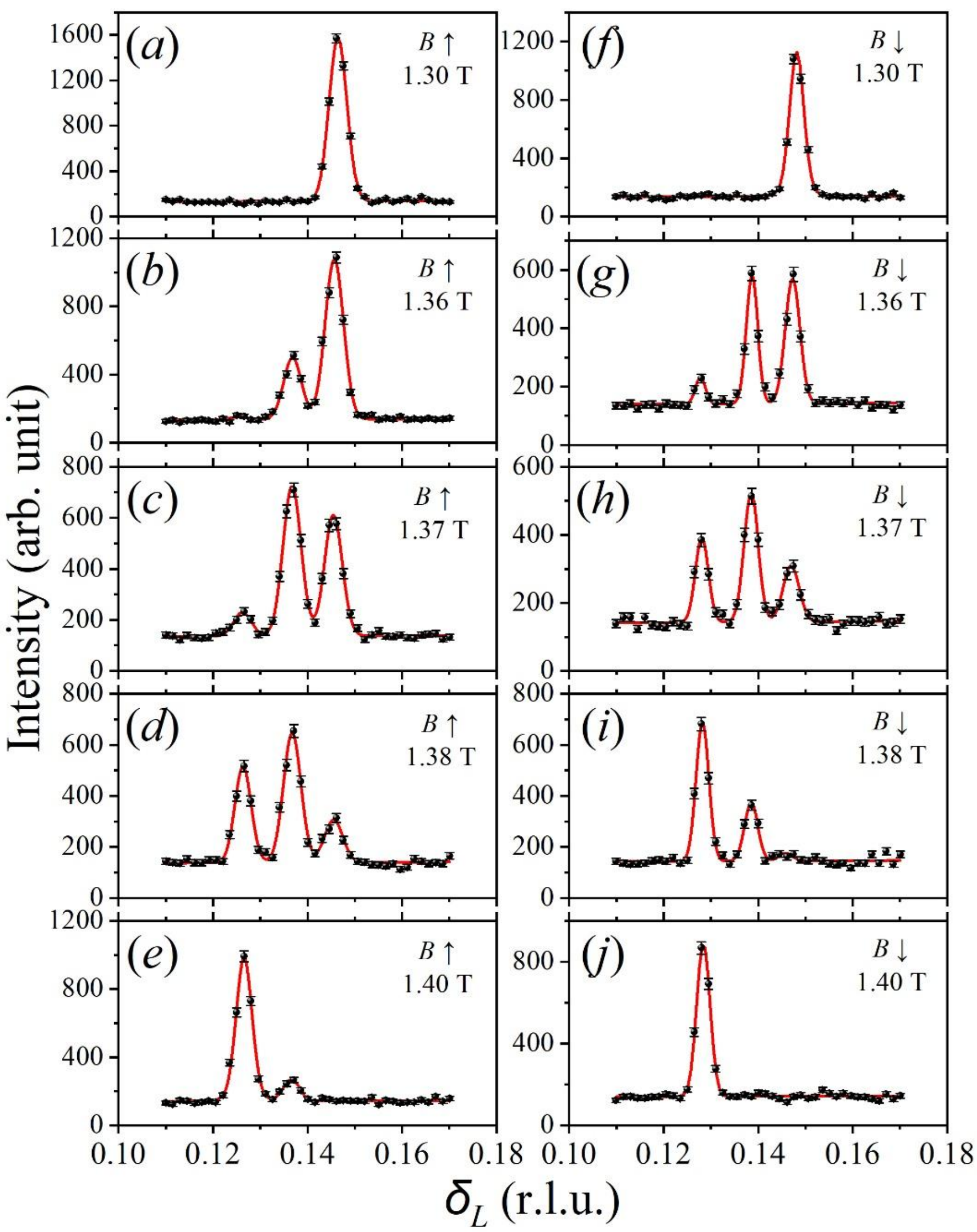

Figure S7: Scattering patterns between 1.3 to 1.4 T at 2 K, where (a) to (e) represent increasing field, and (f) to (j) are for the decreasing field.

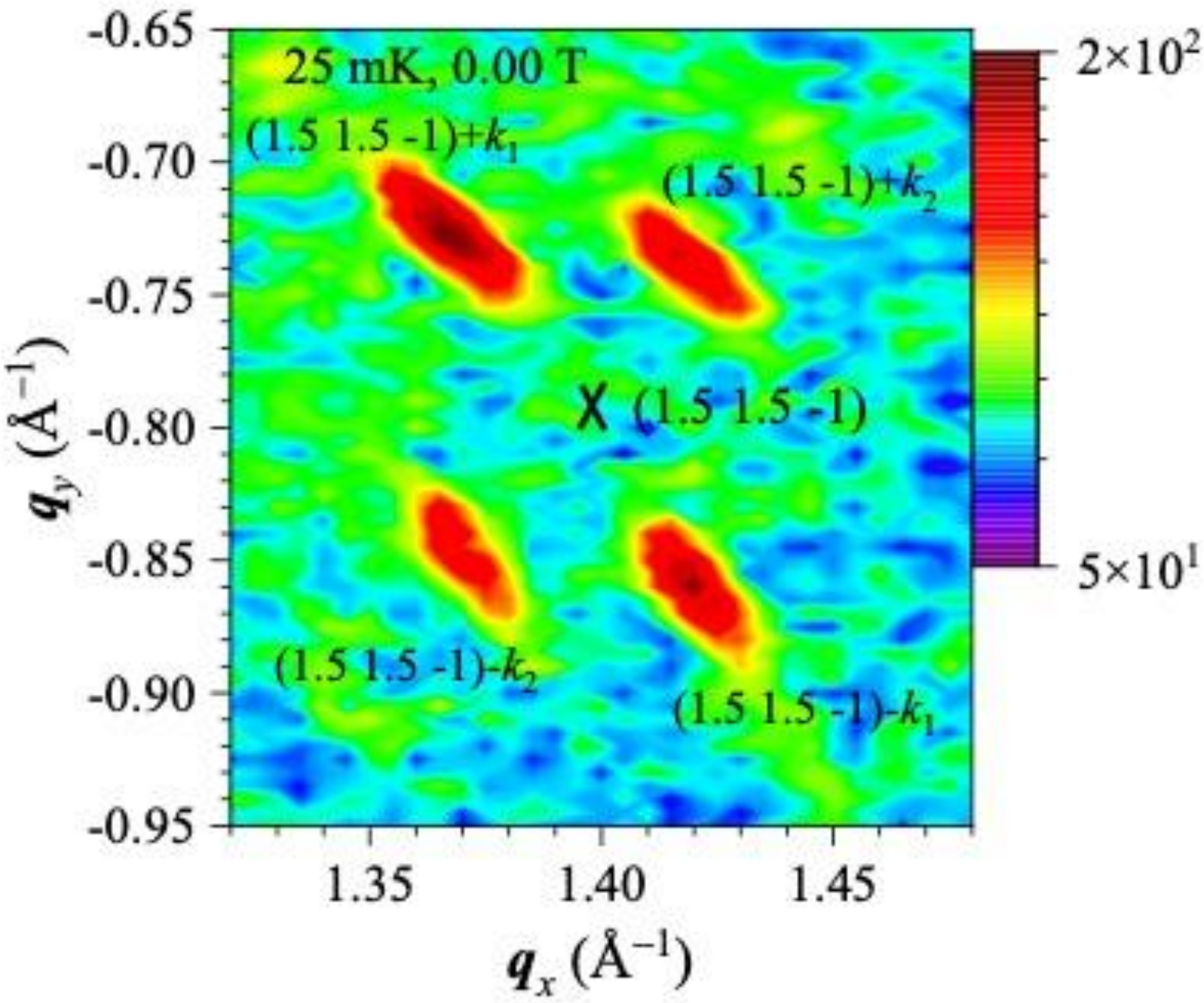

Fig. S8 : A 2D map in $q_x$ *and* $q_y$ which highlights the incommensurate satellites centered around (1.5 1.5 -1) at 25 mK.

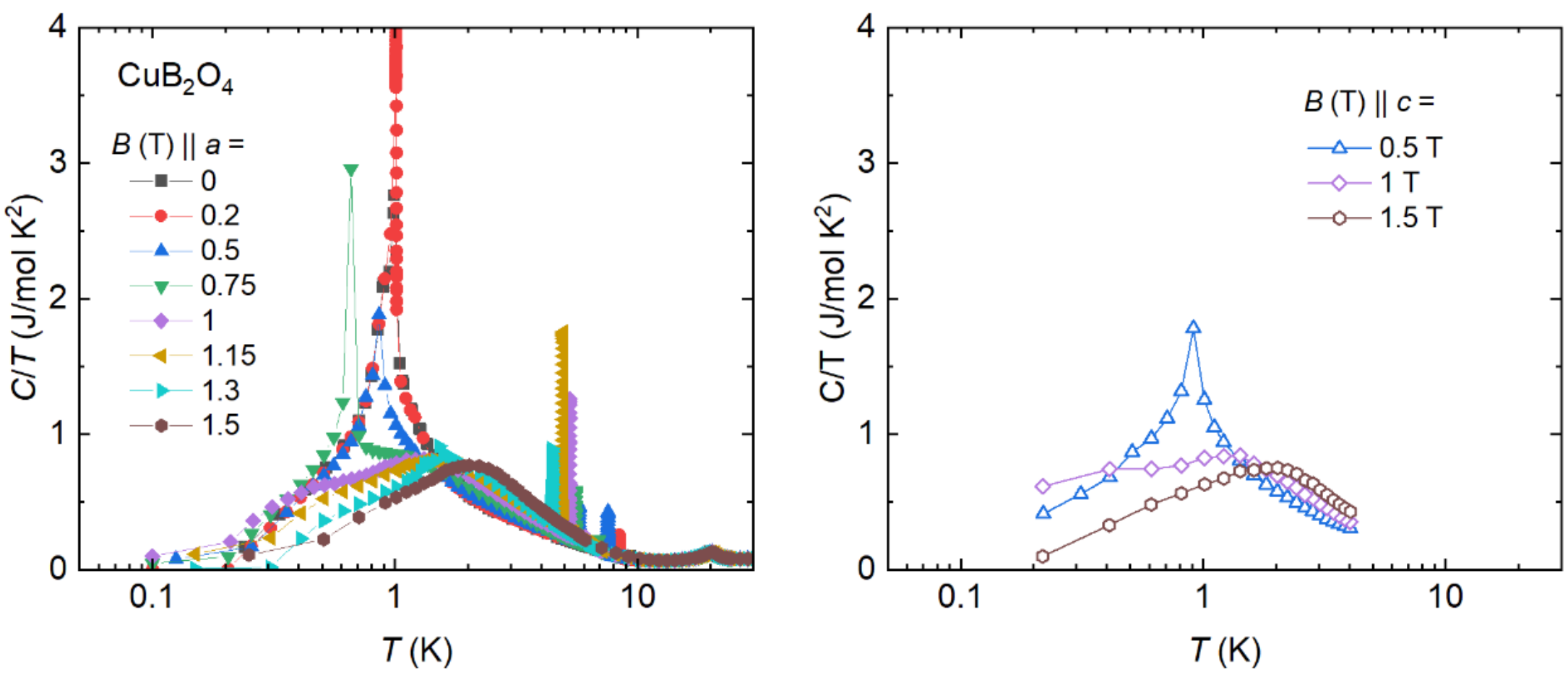

Fig. S9: The specific data with the field applied parallel to the *a*- (left) and *c*-axis (right), respectively.

**AB INITIO STUDY WITH VASP**

In our ab initio study, we use a 6 formula supercell which includes 2 Cu(A) and 4 Cu(B) atoms and a 4×4×4 Monkhorst-Pack $k$-mesh to proceed spinorbit-interaction-included spinor calculations with VASP, where GGA+U density functional (Perdew-Wang 91, and Dudarev's approach for Hubbard U) is adopted. The decision for the $U$ values ($U_{\mathrm{Cu(A)}}$ = 6.4 eV, $U_{\mathrm{Cu(B)}}$=7.0 eV) is made to parameterize the 2 experimental values, one being the valence-conduction band gap, and the other being the gap still above. While our DFT study stays in the commensurate phase, the symmetry of the non-collinear magnet moment pattern is adopted from previous neutron scattering data [9].

As mentioned in the manuscript, the two groups of frontier conduction bands of $CuB_2O_4$ are noticeably flat, the band widths of the groups being 34 meV and 143 meV, as shown in Fig. S10. The flatness leads to two prominent peaks immediately above $E_F$ in the DOS as shown in Fig. S11. From the integrated density plots we depicted two peaks in Fig. S11. The corresponding sources of contributions for those two peaks are identified: The lower energy peak comes from Cu(A) and $O_1$, and the higher energy peak comes from Cu(B) and $O_2 \sim O_4$, illustrated by highlighted plaquettes in the ball-and-stick model in Fig. S11. The plaquette orientations rationalize the orbital characters: for Cu(A), the $z$-related $d$-orbitals barely contribute to the LDOS of the first peak; for Cu(B), the contribution of $d_{xy}$ orbital to the LDOS of the second peak in correspondingly minimal.

The magnetic exchange energies of Cu(A) and Cu(B) are scaled with the band widths. Both plaquette species are isolated. They are not edge- nor corner-shared within each species. At the first glance, that the Cu(B) peak comes from a slightly wider band width seems to imply a stronger coupling among the Cu(B) plaquettes. However, the experimental fact suggests a weaker magnetic ordering of Cu(B) starting at a lower temperature of $\gtrsim$ 8 K, while the Cu(A) magnetic ordering holds at ~ 20 K. The crystal structure model shown in Fig. S12 sheds light on this interesting puzzle. In this case, the Cu(B) plaquette species can be further classified into dimer-like pairs that are normal to $\hat{x}$, and those normal to $\hat{y}$. Judging from geometry, the coupling between these two types of "dimers" should be slim†. Within each type, according to relative distances, the "intra-dimer" coupling is even stronger than the coupling among the Cu(A) plaquettes‡, while the "inter-dimer" coupling is obviously weaker. Therefore, the larger width of the Cu(B) band group in fact comes from the "intra-dimer" coupling, and the weak "inter-dimer" coupling is responsible for Cu(B)'s lower magnetic ordering temperature.

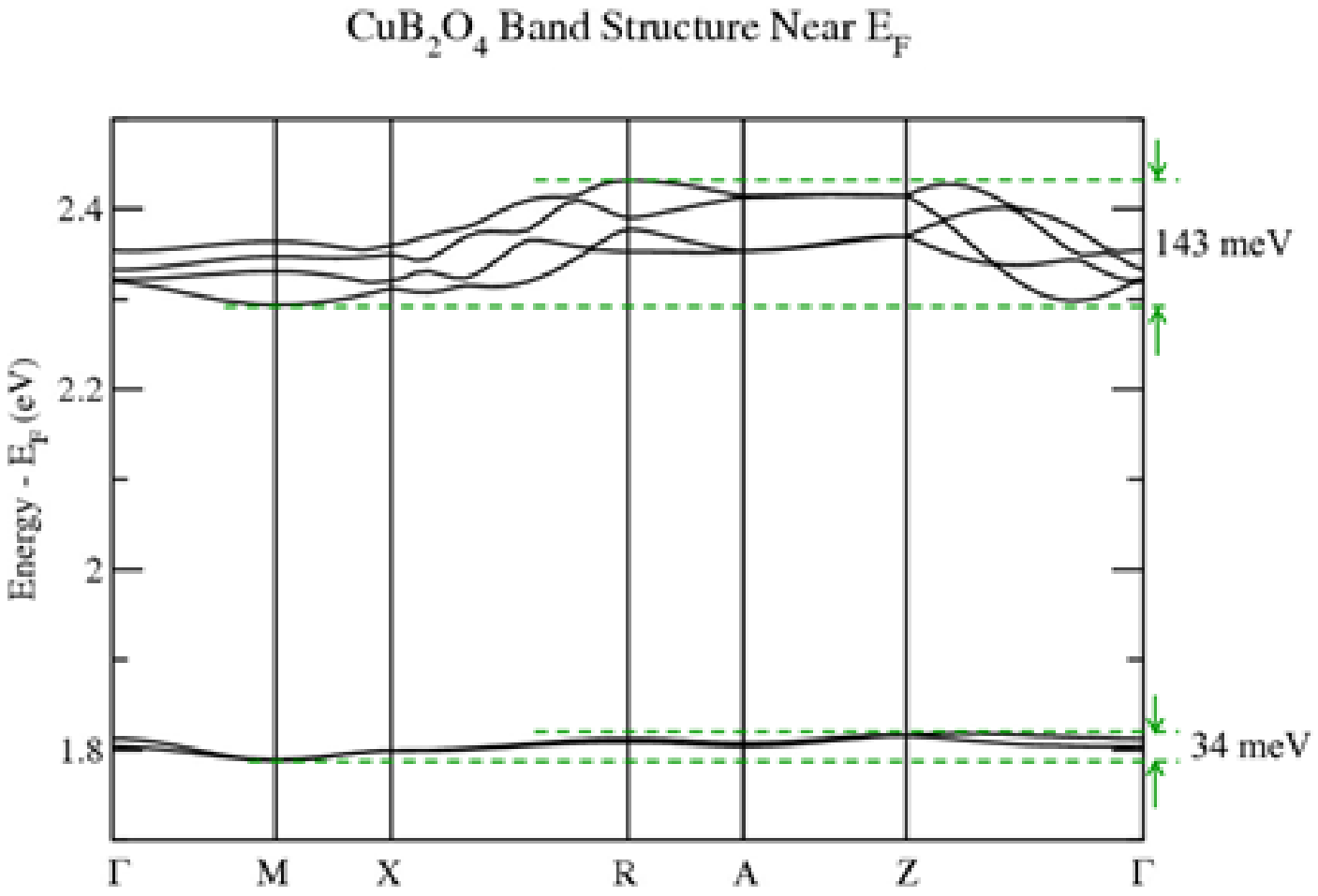

Fig. S10: The energy bands associated with Cu(A) and Cu(B). It is noted that the bandwidth is much narrower than the ordinary band width (at least ~ several eV).

†Although a Cu(B) $\hat{x}$-plaquette's nearest neighbor is a Cu(B) $\hat{y}$-plaquette (of distance 4.51Å, slightly

shorter than that of the two Cu(B) $\hat{x}$-plaquettes with a “dimer-like” pair, which is 4.76 Å), their almost orthogonal orientations and relative position lead to projection that largely weakens the $\sigma$-coupling between them (to $\lesssim 30\%$), which is a smaller scale version of why Cu(A) and Cu(B) plaquettes do not couple.

‡As a reference, the Cu(A)-Cu(A) nearest neighbor distance is 5.92 Å.

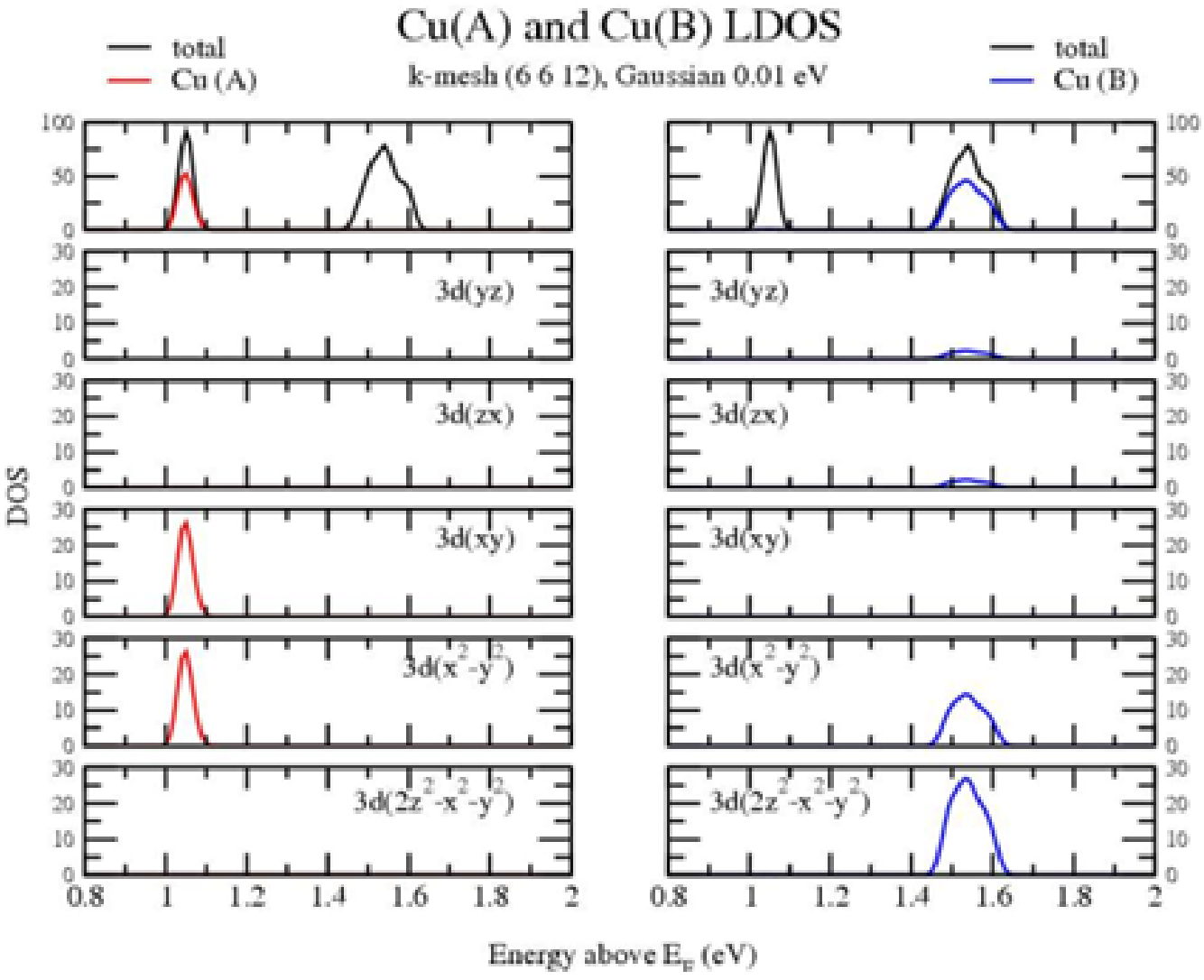

Fig. S11: The charge density integrated over the Cu(A) and Cu(B) band, respectively.

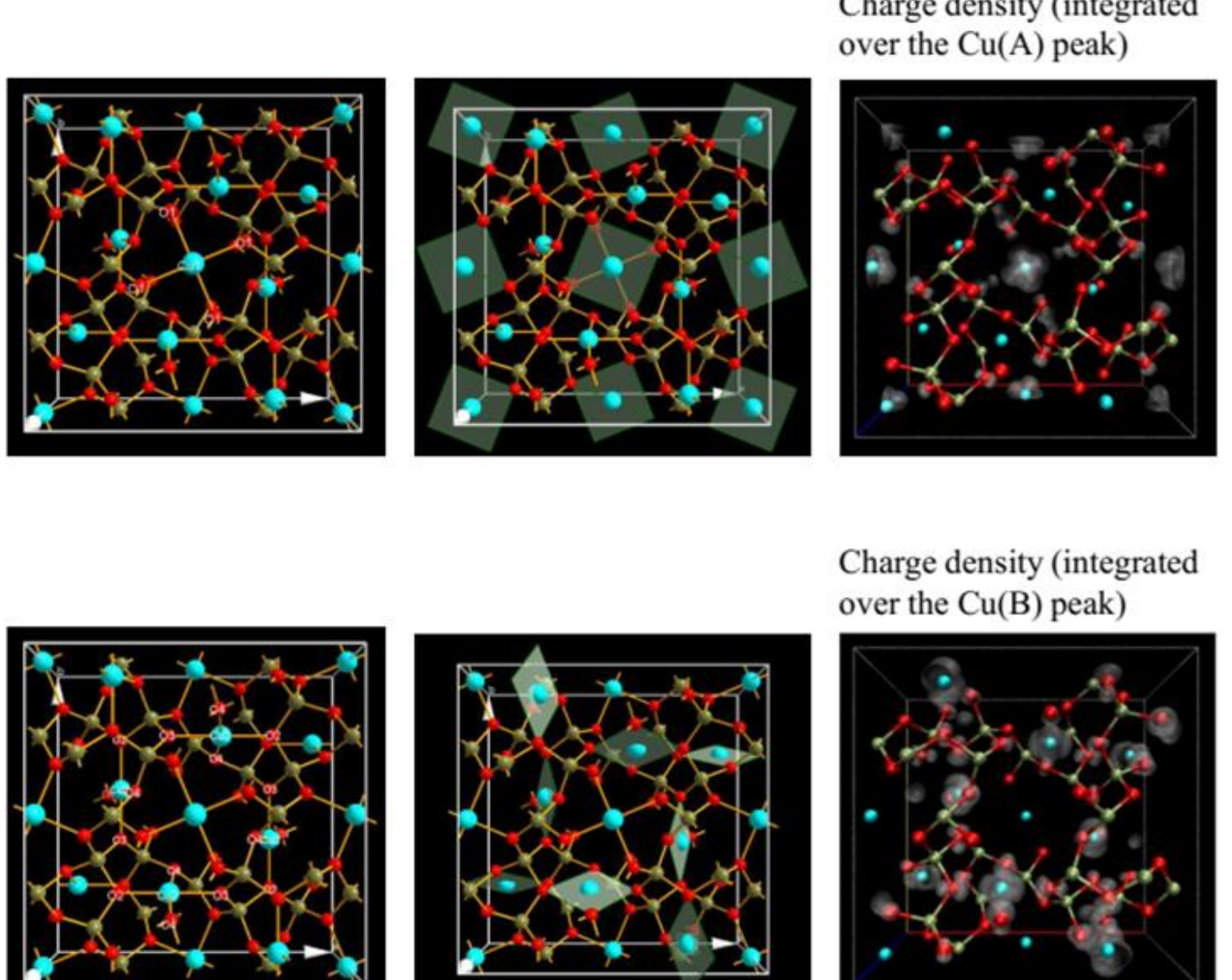

Fig. S12: Upper panels: The Cu(A) plaquettes are composed of Cu(A) and $O_1$ atoms, and normal to (0,0,1). Lower panels: The Cu(B) plaquettes are composed of Cu(B) and $O_2 \sim O_4$ atoms, forming "dimer-like" pairs either facing south-north or east-west. To be precise, within each pair, one plaquette tilts toward $\hat{z}$ by $\sim 20.3°$, while the other tilts toward $-\hat{z}$ by the same angle.